\documentclass[journal=inoraj,manuscript=article]{achemso}

\usepackage{chemformula} 
\usepackage[T1]{fontenc} 

\author{Holger-Dietrich Sa{\ss}nick}
\affiliation{Institute of Physics, Carl-von-Ossietzy Universit{\"a}t Oldenburg, 26129 Oldenburg, Germany}
\author{Fabiana Machado Ferreira De Araujo}
\affiliation{Institute of Physics, Carl-von-Ossietzy Universit{\"a}t Oldenburg, 26129 Oldenburg, Germany}
\altaffiliation{Present Address: Institute of Materials Science, Technische Universit{\"a}t Darmstadt, 64289 Darmstadt, Germany}
\author{Joshua Edzards}
\affiliation{Institute of Physics, Carl-von-Ossietzy Universit{\"a}t Oldenburg, 26129 Oldenburg, Germany}
\author{Caterina Cocchi}
\affiliation{Institute of Physics, Carl-von-Ossietzy Universit{\"a}t Oldenburg, 26129 Oldenburg, Germany}
\alsoaffiliation{Center for Nanoscale Dynamics (CeNaD), Carl-von-Ossietzy Universit{\"a}t Oldenburg, 26129 Oldenburg, Germany}
\email{caterina.cocchi@uni-oldenburg.de}

\title{Impact of Ligand Substitution and Metal Node Exchange in the Electronic Properties of Scandium Terephthalate Frameworks}

\begin{document}
	
	\begin{tocentry}
		
		\includegraphics[height=4.45 cm]{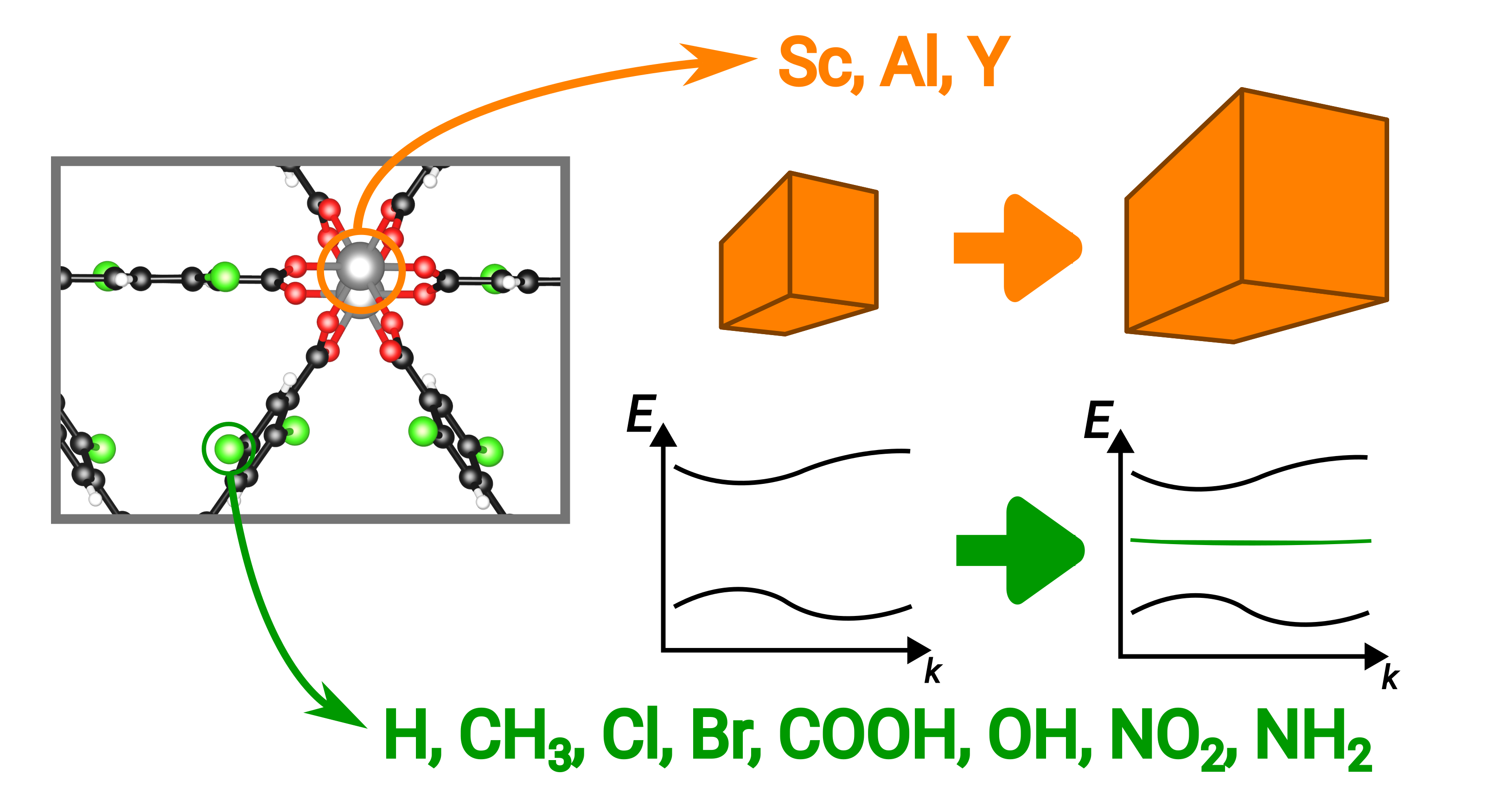}
		
	\end{tocentry}
	
	
	\newpage
	
\begin{abstract}
The search for sustainable alternatives to established materials is a sensitive topic in materials science. Due to their unique structural and physical characteristics, the composition of metal-organic frameworks (MOFs) can be tuned by the exchange of the metal nodes and the functionalization of the organic ligands giving rise to a large configurational space.
Considering the case of scandium terephthalate MOFs and adopting an automatized computational framework based on density-functional theory, we explore the impact of metal substitution with the earth-abundant isoelectronic elements Al and Y, and of ligand functionalization of varying electronegativity.
We find that structural properties are strongly impacted by the metal ion substitution and only moderately by ligand functionalization. In contrast, the energetic stability, the charge density distribution, and the electronic properties -- including the size of the band gap -- are primarily affected by the termination of the linker molecules. Functional groups such as OH and NH$_2$ lead to particularly stable structures thanks to the formation of hydrogen bonds and affect the electronic structure of the MOFs by introducing mid-gap states. 
\end{abstract}

	\newpage
	\section{Introduction}
	
	
	
	Metal-organic frameworks (MOFs) are porous materials formed by metal atoms bound together by organic linkers~\cite{rows-yagh04mmm,kit+04acie}.
	The peculiar structure and chemical tunability of MOFs~\cite{kala-cohe20acscs,syed+20acscat} offer great potential in many technological areas including gas storage and conversion~\cite{devi+12jmc,sned+14aem,alez+15jacs}, optoelectronics~\cite{khaj+11jpcc,cui+12cr,yang+19cc}, and catalysis~\cite{lee+09csr,zhan+18poly}.
	The size of the pores, as well as their electronic and optical properties, can be modulated by the choice of the metallic nodes and/or the molecular ligands~\cite{zhan+19cec,kim+21cec}. 
	The latter, furthermore, can be functionalized with specific groups having electron-withdrawing or -donating ability, thus offering an additional handle to tune the characteristics of the MOFs~\cite{lin+09jacs,lim+22rsca}.
	It is evident that so many degrees of freedom give rise to an enormous configurational space that calls for high-throughput (HT) screening approaches to be properly explored.
	
	Experimentally, HT synthesis of MOFs has been established since the end of the past century~\cite{klei+98acie} and has been exploited, among other purposes~\cite{stoc-bisw12cr}, to maximize the performance of zeolitic imidazolate frameworks for \ch{CO2} capture~\cite{bane+08sci}, and to optimize the structure of porous chromium terephthalate to host particularly large guest molecules~\cite{fere+05sci}.
	More recently, the development of computational HT methods based on density-functional theory (DFT)~\cite{pizzi+16cms,jain+11cms} has opened up the opportunity to design MOFs \textit{in silico}~\cite{rose+19jcc,isla+23npjcm,zhan+23acsami}.
	The advantages of this approach are numerous: It does not demand experimental synthesis and characterization; it enables exploring a potentially infinite amount of constituent combinations; it offers an overview of the fundamental properties of the MOFs on a quantum-mechanical level. 
	
	The current quest for new materials to adhere to sustainability requirements further stimulates HT computational studies on MOFs with the task of identifying suitable alternatives to specific ligand molecules that are toxic or hazardous~\cite{wisn+23cej}.
	Likewise, many metallic species pose challenges regarding availability and extraction costs.
	Scandium is a prominent example in this regard.
	While being presently in high demand due to the favorable mechanical properties of Al-Sc alloys~\cite{venk+04msea,roys-ryum05imr} and for its applicability in medical laser technologies~\cite{scho+07jada}, this element is tremendously hard to harvest.
	Currently, scandium is mainly recovered as a byproduct from the production of other metals~\cite{wang+11metallurgical} or from bauxite residues~\cite{zhan+16rm}.
	Furthermore, its production is concentrated in a few world areas, which do not include those mostly requesting it, such as Europe~\cite{vand+23}.
	Scandium-based MOFs have recently emerged as promising materials for carbon oxide sequestration~\cite{mowa+11ic,gree+14acie,pill+15jpcc} and fluorescence sensing~\cite{xie+19cs,zhan+20dt}.
	However, the limited availability of Sc calls for alternatives.
	Aluminum and yttrium, both isoelectronic with scandium and more abundant on the Earth's crust, are seen as suitable substitutes for this element in MOFs.
Also, a recent study~\cite{pras+19jmca} has shown that Al replacement of Sc ions in a Sc-based MOF enhances the \ch{CO2} adsorption ability of the material.
	
	Motivated by this experimental evidence and by the interest in discovering sustainable alternatives to established MOFs, we present a computational study based on DFT investigating the structural and electronic properties of scandium terephthalate scaffolds modified by metal-ion substitution and ligand functionalization.
	By applying an in-house implemented automated workflow for \textit{ab initio} calculations~\cite{sass-cocc22jcp}, we construct 24 structures and analyze their equilibrium geometries and their energetic stability at varying metal nodes and molecular functionalization.
 We discuss the larger impact of the metallic species on the structural properties in contrast with the moderate effect of the ligand substituents which merely induce some steric hindrance.
Interestingly, all explored MOFs are stable and linker functionalization has a large impact in this respect.
	By means of partial charge analysis, we shed light on the bonding among the involved species revealing also in this case the significant influence of ligand terminations.
 We finally discuss the electronic properties of the considered systems, which are all large-band-gap semiconductors expected to absorb ultraviolet radiation.
 The functional groups influence considerably the size of the fundamental gap.
 In particular, OH and \ch{NH2} groups give rise to mid-gap states that alter even qualitatively the electronic characteristics of the MOFs.

	\section{Methodology}
	
\begin{figure}
		\centering
		\includegraphics[width=0.5\textwidth]{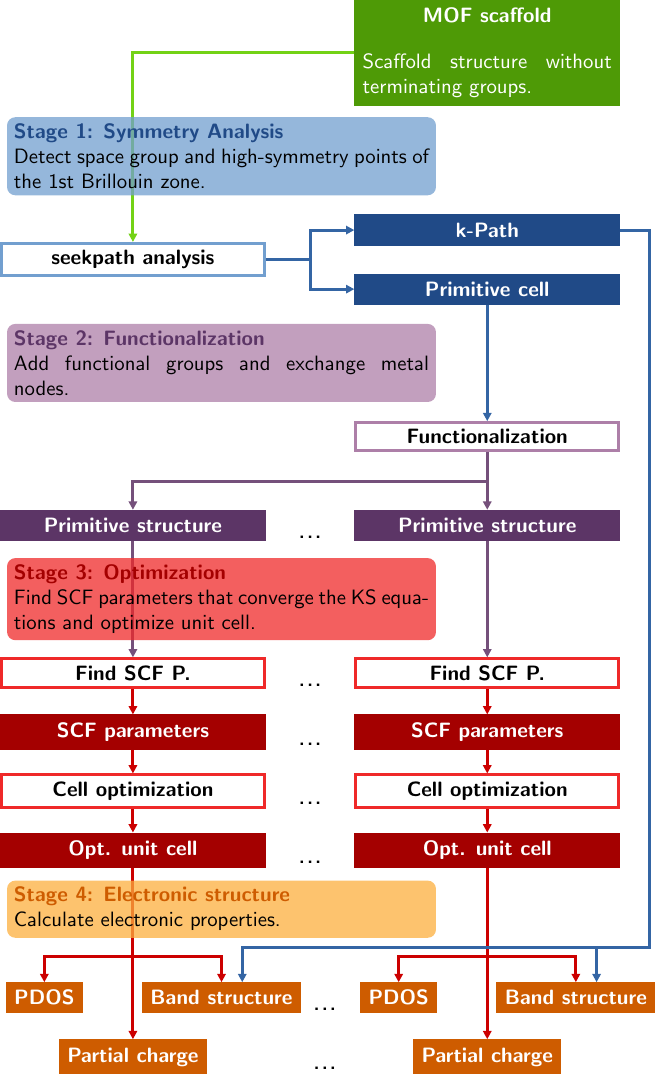}
		\caption{
        Sketch of the workflow adopted in this work supported by the AiiDA infrastructure. In the first stage, the MOF structure is loaded via a CIF file or directly from an online database. The python module \texttt{seekpath} analyzes the high-symmetry points of the first Brillouin zone and detects the space group of the structure outputting the primitive cell and the k-path. In the second step, the primitive cell undergoes the desired structural modifications (metal node exchange and ligand functionalization). In step 3, each structure is optimized until the convergence threshold is reached. Finally, the electronic properties including band structure, PDOS, and partial charges are calculated.
        }
		\label{fgr:workflow}
	\end{figure}

	The computational workflow used in this study is implemented in an in-house developed library embedding routines for data mining, HT DFT calculations based on the AiiDA infrastructure~\cite{pizzi+16cms, hube+20sd}, and post-processing tools.
	This package, initially designed for inorganic crystals~\cite{sass-cocc22jcp}, has been purposely tailored here to investigate MOFs (see Figure~\ref{fgr:workflow}).
The initial input includes structural information about the scaffold which, in this case, is scandium terephthalate.
In the first computational step, the primitive unit cell of the constructed structures, their space group, and k-paths are identified using the python libraries \texttt{seekpath}~\cite{hinu+17cms} and \texttt{spglib}~\cite{togo-tana18arxiv}.
The backbones of the organic ligands are stripped off of their native terminations and subsequently equipped with the chosen functional groups.
Likewise, the Sc nodes are replaced with Y and Al atoms, giving rise to the final pool of input structures for the automatized DFT calculations.
The remaining part of the workflow is equivalent to the one presented in Ref.~\citenum{sass-cocc22jcp}, to which we redirect interested readers for further information. The specific details of the DFT runs are optimized for MOFs. Specifically, the threshold for the minimization of interatomic forces is set to 0.025~eV/{\AA}, a relatively large parameter for stiff and dense inorganic crystals but suitable for flexible and porous frameworks; for the same reason, during optimization, only the angles of the unit cell are constrained instead of the space group; finally, the k-mesh is constructed with equidistant points separated by 0.2~{\AA}$^{-1}$, which is adequate to accurately sample the relatively small Brillouin zones of the considered MOFs.

All DFT calculations are performed with the code CP2K~\cite{kueh+20jcp} which implements the Gaussian and plane-wave method~\cite{vand+05cpc}.
	Core electrons are accounted for by the dual-space pseudopotentials of the Goedecker-Teter-Hutter type~\cite{goed+96prb} while valence electrons are represented within the MOLOPT triple-$\zeta$ basis set including two polarization functions shipped with the code.
	To ensure numerically converged results, the plane-wave cut-off and the relative cut-off values are set to 600~Ry and 100~Ry, respectively.
	The Perdew-Burke-Ernzerhof (PBE) functional~\cite{perd+96prl} is used in all calculations in conjunction with Grimme D3 method~\cite{grim+10jcp} to account for long-range dispersion interactions.
	The code \texttt{critic2}~\cite{oter+14cpc} is employed to calculate partial charges within the Bader method~\cite{laid-bade90jcp} and the Yu-Trinkle integration scheme~\cite{yu-trin11jcp}.

	\section{Results and Discussion}

	\subsection{Structural Properties and Stability}
	
	\begin{figure}
		\centering
		\includegraphics[width=\textwidth]{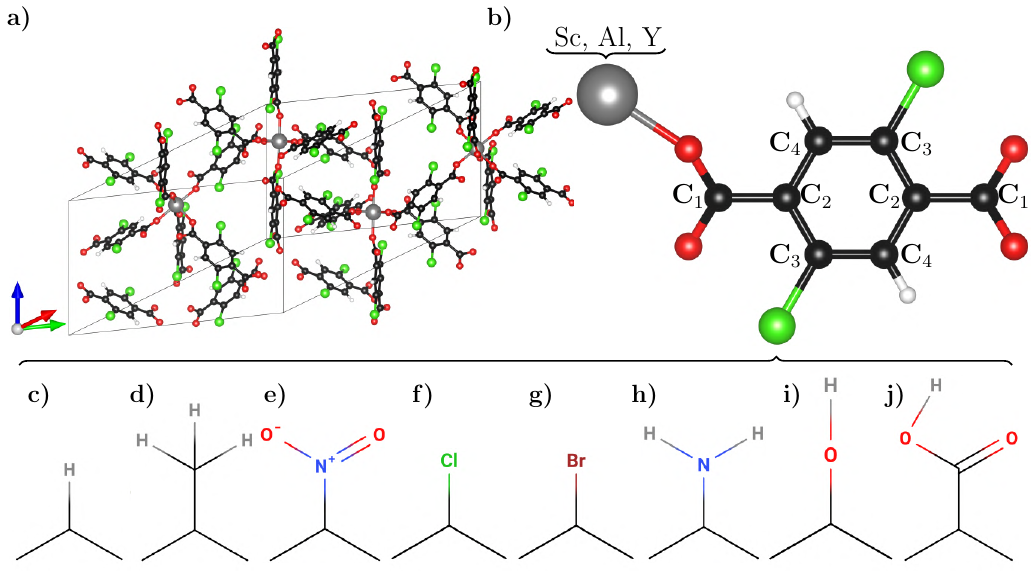}
		\caption{a) Ball-and-stick representation of the orthorhombic primitive unit cell of the scandium terephthalate Sc$_2$(BDC)$_3$ plotted with VESTA~\cite{VESTA}. b) Building unit of Sc$_2$(BDC)$_3$ with C atoms depicted in black and indexed according to their (in)equivalent sites, H atoms shown in white, O atoms in red, and the metal center in gray: Al and Y are considered in addition to Sc. The adopted functional groups, marked in green in panels a) and b), include c) H, d) \ch{CH3}, e) \ch{NO2}, f) Cl and g) Br, h) \ch{NH2}, i) \ch{OH}, and j) \ch{COOH}. }
		\label{fgr:mof_crystal}
	\end{figure}
	
The MOF studied in this work is the small-pore scandium terephthalate with chemical formula \ch{Sc2(BDC)3}, consisting of Sc metal nodes and of 1,4-benzene-dicarboxylate as molecular linkers, see Figures~\ref{fgr:mof_crystal}a,b.
The pores with a radius of about 3~\AA{} favor gas absorption~\cite{mowa+11ic}.
This MOF undergoes a phase transition at 225~K from the monoclinic (space group $C2/c$) to the orthorhombic phase (space group $Fddd$, see Figure~\ref{fgr:mof_crystal}a) with the latter exhibiting negative thermal expansion~\cite{mowa+11ic}.
Note that this characteristic is common to other Sc-based materials such as \ch{ScF3}~\cite{grev+10jacs,laza+15prb,oba+19prm}.
In this study, the orthorhombic phase of scandium terephthalate extracted from Ref.~\citenum{mowa+11ic} is used as a basis to construct the substituted MOFs.
Sc is replaced by the isoelectronic elements Al and Y, and the linker molecules are functionalized at the sites marked in green in Figure~\ref{fgr:mof_crystal}b.
In addition to the H termination (Figure~\ref{fgr:mof_crystal}c), methyl, nitrogen dioxide, atomic Cl and Br, the amino group, the hydroxyl group, and the carboxylic group are considered (see Figures~\ref{fgr:mof_crystal}d-j, respectively).
This way, 24 different structures are obtained as an input for the DFT calculations.

\begin{figure}
	\centering
	\includegraphics[width=0.5\textwidth]{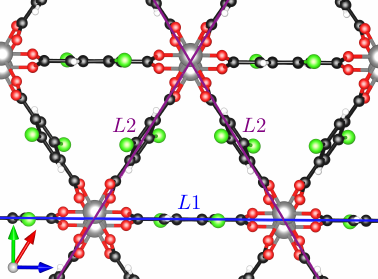}
\caption{Inequivalent sites, $L1$ and $L2$, of the linker molecules of the scandium terephthalate MOFs.}		
\label{fgr:l1_l2_linkers}
\end{figure}

Regardless of the specific composition, the topology of the considered MOFs is characterized by two inequivalent linker molecules.
The corresponding sites are labeled herein as $L1$ and $L2$ and appear in the structure with a ratio 1:2 (see Figure~\ref{fgr:l1_l2_linkers}). 
Molecules on the $L1$ site lie parallel to one of the crystal axes while those in $L2$ form an angle of approximately 60$^{\circ}$ with it, giving rise to the peculiar triangular shape of the pore in this MOF.
For the linkers in $L1$, the functional groups lie on the same plane of the phenyl ring, thereby inducing a torsion in the \ch{CO2} groups binding the ligands to the metal nodes.
In contrast, in $L2$, the functionalized carbon rings as well as the \ch{CO2} groups binding them to the metal atoms are slightly twisted (see Figure~\ref{fgr:l1_l2_linkers}).
These qualitative differences can be quantified by evaluating the distances between the metal ion and the O atom of the \ch{CO2} groups and by the dihedral angle of the latter; these results are reported in the Supporting Information (SI) on Tables~S1 and S2. 
 
\begin{figure}
\centering
\includegraphics[width=0.5\textwidth]{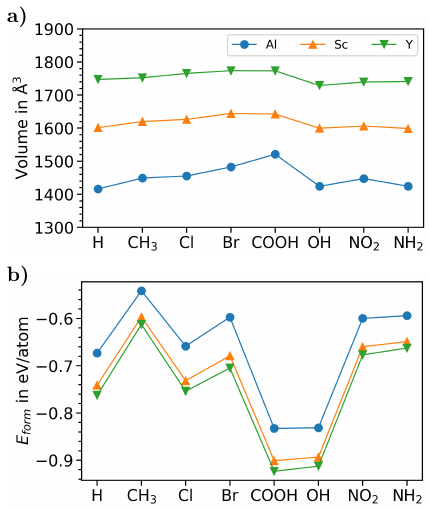}
\caption{a) Optimized unit-cell volumes and b) formation energies ($E_{Form}$) of all considered MOFs.}
\label{fgr:vol_form}
\end{figure}

After these considerations, we are equipped for the analysis of the structural characteristics of the relaxed unit cells of the considered MOFs which we assess in terms of their volume.
As shown in Figure~\ref{fgr:vol_form}a (the raw data are reported in Table~S3), the size of the unit cell varies significantly depending on the metal nodes while ligand functionalizations introduce changes on a smaller scale.
This behavior can be explained by the different atomic radii of the considered metal atoms: the largest (smallest) volumes pertain to the structures with Y (Al) atoms, which indeed have the largest (smallest) size among the species adopted for the nodes.
Similar trends are found also for the distances between the metal atoms and the oxygen atoms belonging to the \ch{CO2} groups of the linker, see Table~S1.

Focusing now on the effects induced in the volume by ligand functionalization, we notice similar but not identical trends for the three considered scaffolds (see Figure~\ref{fgr:vol_form}a).
In the Al-based MOFs, the smallest volume is found for the H-passivated linker while all functionalizations lead to an increase in the unit-cell size.
This finding can be explained in terms of steric hindrance: with COOH this effect is particularly pronounced and can be intuitively understood considering the large size of this group. 
In the Sc- and Y-based MOFs, we notice some slightly different trends. 
Keeping the H-passivated structure as a reference, we notice that only methyl functionalization and the highly electron-withdrawing terminations \ch{Cl}, \ch{Br}, and \ch{NO2} lead to a larger unit-cell volume.
With OH and \ch{NH2}, instead, the Sc- and Y-based MOFs experience a slight decrease in volume due to the formation of hydrogen bonds between the oxygen atoms of the BDC and the H atom of the functional group.
Corresponding interatomic separations range between 1.952--2.019 with OH and 1.720--1.778~\AA{} with \ch{NH2}, see Table~S4.
It is worth noting that hydrogen bonds are formed also in the presence of the \ch{COOH} group but only at the $L2$ site.
In the Al-based MOFs, characterized by the smallest volumes, their effect is dramatic and leads to symmetry breaking.
	
After the examination of the structural properties, we now move to the analysis of the stability which we assess by examining the formation energy per atom.
This quantity is computed as the difference between the total energies of the MOFs and the most stable crystalline phases of its constituting elements taken from Materials Project~\cite{jain+13aplm}:
	\begin{equation}
		\begin{split}
			E_{form} = E(MOF) &- \dfrac{2}{44+6\, n} E(\text{M}) - \dfrac{12}{44+6\, n} E(\text{O}) - \dfrac{24}{44+6\, n} E(\text{C}) \\
			&- \dfrac{6}{44+6\, n} E(\text{H}) - \dfrac{6}{44+6\, n}\sum_{i=1}^{n} E(\text{A}^{\ch{X}}_i).
			\label{eq:form}
		\end{split}
	\end{equation}		
 In Eq.~\eqref{eq:form}, $E(MOF)$ is the total energy per atom of the relaxed MOF while
	$E(\text{M})$, $E(\text{O})$, $E(\text{C})$, $E(\text{H})$ and $E(\text{A}^{\ch{FG}}_i)$ are the total energies per atom of the elemental crystalline phases of the metal atoms (M~=~Sc, Al, Y), of oxygen, carbon, hydrogen, and of each atom $A^{\ch{X}}_i$ of the functional group (X), respectively.
	Under these conditions, zero-point energies and thermal contributions are not included.

 As reported in Figure~\ref{fgr:vol_form}b (see also Table~S6), we find similar trends of stability for MOFs with the same metal node.
	The Al-based frameworks, which are characterized by the smallest volumes (Figure~\ref{fgr:vol_form}a), have the least negative formation energies, namely, they are less stable than their Sc- and Y-based siblings.
	Conversely, the large-volume Y-containing MOFs feature the most negative formation energies suggesting their larger stability over the other considered scaffolds.
	Functionalization with Cl and Br atoms as well as with the \ch{NH2}, \ch{NO2}, and \ch{CH3} groups gives rise to less stable structures compared to those containing OH and COOH, which stabilize the MOFs through the formation of hydrogen bonds (see Table~S4).
	Although the \ch{NH2}-group forms hydrogen bonds as well, the corresponding bond lengths are larger and therefore contribute less substantially to stabilizing the MOFs.
	
\subsection{Partial Charge Analysis}

\begin{figure}[h!]
    \centering
    \includegraphics[width=\textwidth]{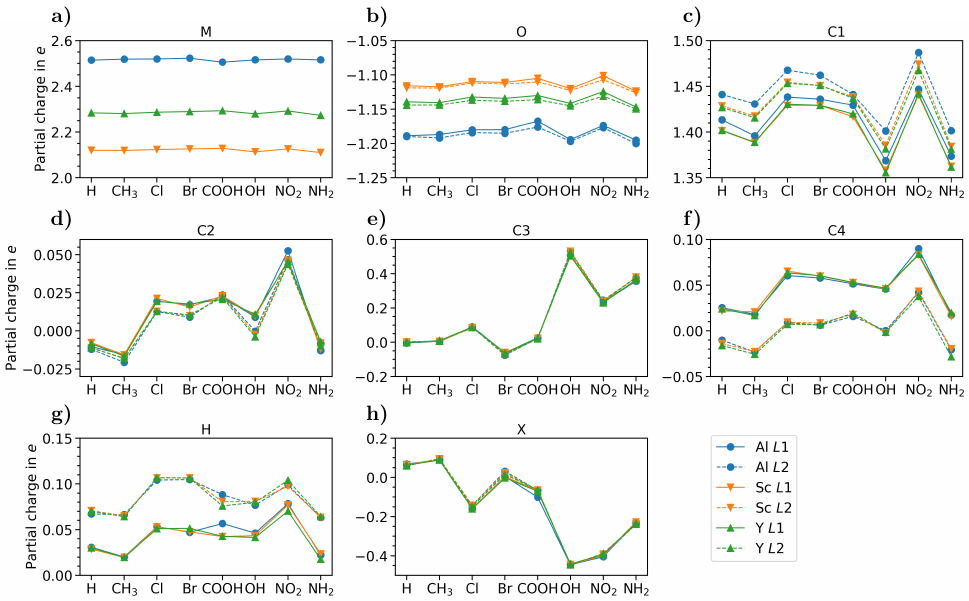}
    \caption{Partial charges calculated with the Bader scheme of the a) metal ions (M), b) O atoms, c)-f) C atoms, g) H atoms in the backbone of the ligand, and h) functional group as a whole in all the considered MOFs. Solid and dashed lines in panels b)-h) denote the molecular $L1$ and $L2$ sites, respectively, while the metal species in the scaffold are indicated by blue circles (Al), down-pointing orange triangles (Sc), and up-pointing green triangles (Y).
    }
    \label{fgr:partial_charges}
\end{figure}
	
We now turn to the analysis of the partial charges of the considered MOFs calculated using the Bader scheme~\cite{laid-bade90jcp}.
We examine the results obtained for the metal atoms (Figure~\ref{fgr:partial_charges}a) as well as for all the species included in the backbone of the linkers (Figures~\ref{fgr:partial_charges}b-g) and in their functionalization (Figure~\ref{fgr:partial_charges}h).
In this analysis, we distinguish between the values obtained for the ligands at the inequivalent sites $L1$ and $L2$.
 
The partial charges calculated for the metal atoms are positive in all structures, with the largest values pertaining to Al and the lowest to Sc (see Figure~\ref{fgr:partial_charges}a).
This trend may be puzzling considering the electronegativity of these elements.
 However, in the MOF environment, the metal atoms form ionic bonds with the neighboring oxygens, as testified by the results plotted in Figure~\ref{fgr:partial_charges}b.
It can also be noted in passing that in a highly electronegative environment Sc atoms exhibit even lower partial charges than those shown in Figure~\ref{fgr:partial_charges}a, as recently discussed in a first-principles study on \ch{ScF3}~\cite{mach+23ic}.
Considering the values displayed in Figure~\ref{fgr:partial_charges}a,b (see also Tables S7--S9), it is evident that the positive charge on the metal ions is almost entirely compensated by the negative charge of the oxygens bound to them.
Analyzing the trends according to the ligand functionalization, we find that the ionicity of the metal-oxygen bond is more pronounced in the MOFs hosting \ch{OH} and \ch{NH2} groups.
This behavior can be understood by recalling that in the presence of these terminations, hydrogen bonds are formed: both \ch{OH} and \ch{NH2} make available an additional hydrogen atom in the vicinity of the oxygens thus enhancing the partial charges of the latter.
This hypothesis is confirmed by considering the results obtained for the carbon atom C1, which is bound to both O atoms in the BDC linker (Figure~\ref{fgr:mof_crystal}b), and indeed, is positively charged in all considered MOFs (Figure~\ref{fgr:partial_charges}c).
The smallest values for the partial charges are obtained in the structures functionalized with \ch{OH} and \ch{NH2} groups, while the largest ones are collected for the structures with \ch{NO2} ligand termination, again mirroring the trend seen in Figure~\ref{fgr:partial_charges}b for the O atoms.

We continue this analysis by inspecting the partial charges on the C atoms of the phenyl ring (Figures~\ref{fgr:partial_charges}d-f) as well as of their H and X terminations (Figures~\ref{fgr:partial_charges}g-h), with H being the hydrogen passivating the rings of all systems (Figure~\ref{fgr:mof_crystal}b) and X indicating the varying functional atoms or groups.
At a glance, it is evident that the electronic distribution in those species is negligibly influenced by the metal node.
C2, C4, and H atoms show very small partial charges, thereby reflecting the covalent character of their bonds.
However, especially for C4 and H, the values obtained in the molecules at the $L1$ and $L2$ sites differ visibly, while no significant differences are noticed for C2.
These findings can be understood considering the location of the corresponding atoms in the linker molecule (see Figure~\ref{fgr:mof_crystal}b): C4 is passivated by H while C2 is surrounded only by carbon atoms. 
Both C4 and H are generally characterized by positive charges, which become negative in C4 atoms at the $L2$ molecular site in the presence of the electron-donating functionalizations such as H, \ch{CH3}, and \ch{NH2} (see Figure~\ref{fgr:partial_charges}e and \ref{fgr:partial_charges}h).
In contrast, with \ch{NO2} terminations, the partial charges of C4 undergo a visible increase in both $L1$ and $L2$ ligands due to the electron-withdrawing nature of this group.
It should be noted, though, that the absolute values for the charges on C4 remain below 0.1 electrons (see Figure~\ref{fgr:partial_charges}f).
Finally, for C3 and X, which are bound to each other (see Figure~\ref{fgr:mof_crystal}b), we notice mirror trends. 
The large positive charges ($>$0.25~e$^-$) accumulated on C3 with X = \ch{OH}, \ch{NO2}, \ch{NH2} (Figure~\ref{fgr:partial_charges}e) are reflected by the equally large but negative values on the functional groups (Figure~\ref{fgr:partial_charges}h).
Conversely. with the other ligand terminations (H, \ch{CH3}, Br, and \ch{COOH}), we obtain very small partial charges for C3.

\subsection{Electronic Properties}
	
\begin{figure}
	\centering
	\includegraphics[width=0.5\textwidth]{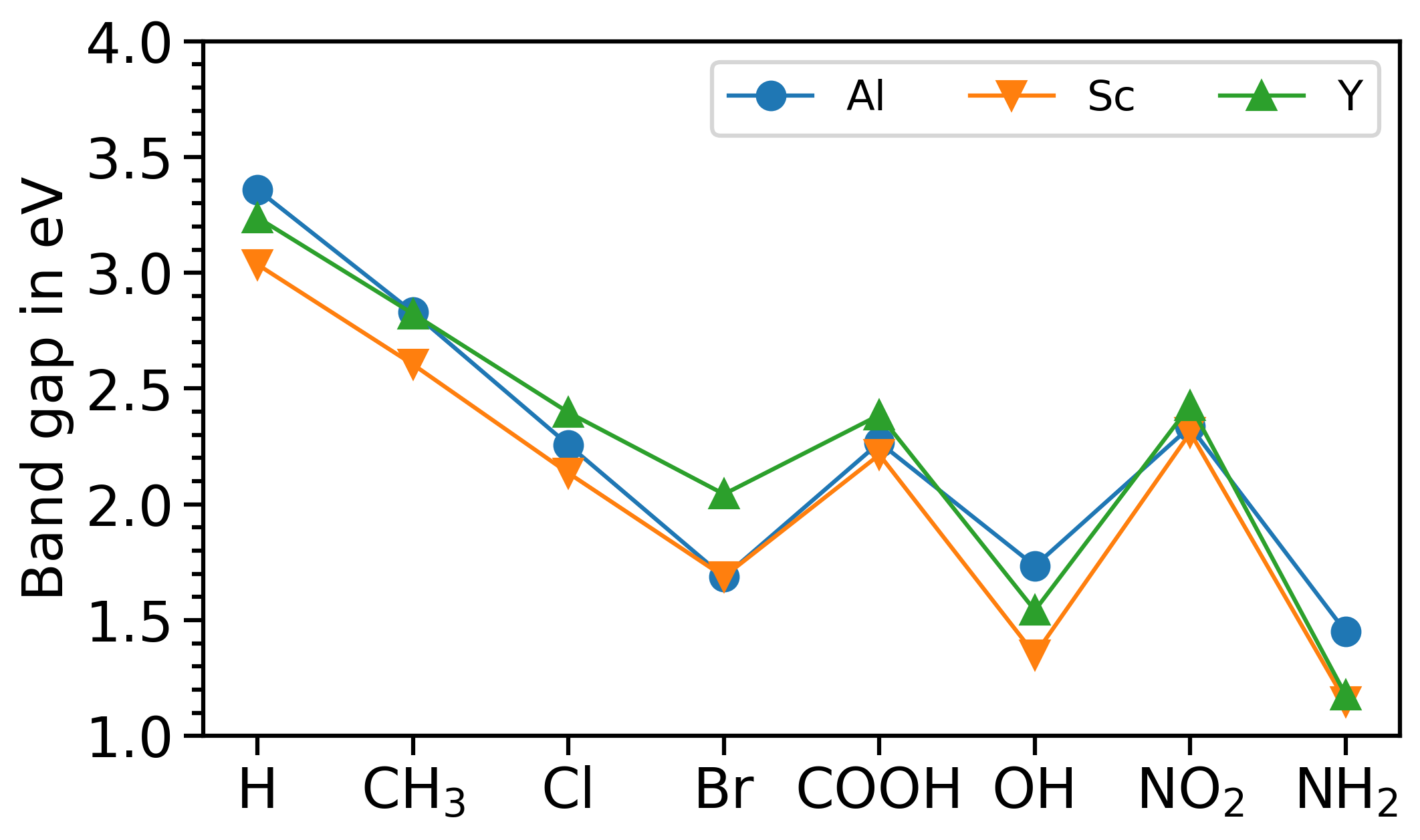}
	\caption{Calculated band gaps of all the considered MOFs.}
	\label{fgr:band_gaps}
\end{figure}
	
We start the analysis of the electronic properties by looking at the band gaps of all the considered MOFs.
By inspecting Figure~\ref{fgr:band_gaps} (see also Table~S6), we notice at a glance that variations induced by linker functionalization are on the order of an eV, while the exchange of the metal node leads to fluctuations of hundreds of meV at the most.
The MOFs with H-passivated linkers exhibit the largest band gaps, ranging between approximately 3.00~eV with Sc up to 3.36~eV with Al.
All the other considered functional groups induce a decrease in the band gap.
With \ch{CH3} termination, the obtained values for the gaps range from 2.60~eV with Sc to 2.82~eV and 2.83~eV with Al and Y nodes, respectively. 
The gaps computed with the linker functionalizations Cl, COOH, and \ch{NO2} are quite similar, being in the range 2.13--2.31~eV with Sc, 2.26--2.34~eV with Al, and 2.39--2.43~eV with Y. 
It is worth noting that in the presence of \ch{NO2} termination, the influence of the specific metal node on the gap is less than 150~meV.
With Br bound to the linker molecules, the MOFs with Al and Sc nodes are predicted to have a band gap of 1.69~eV while with Y it increases to 2.04~eV. 
On the other hand, with OH and \ch{NH2}, the obtained band-gap values are significantly smaller: with the former functionalization, the results range from 1.35~eV with Sc to 1.73~eV with Al, while with \ch{NH2}, the structures including Sc and Y nodes feature a gap of 1.15~eV and 1.18~eV, respectively, whereas the one with Al has a gap of 1.45~eV.

It should be stressed that the band-gap values plotted in Figure~\ref{fgr:band_gaps} are obtained with the semi-local functional PBE, which is known to underestimate this property in crystalline materials up to 50\% of their actual values~\cite{mart20boook}.
On the other hand, qualitatively, the trends provided by PBE are usually reliable and Figure~\ref{fgr:band_gaps} should be interpreted as such.
This assessment is supported by the fact that our trends for the band-gaps are consistent with earlier studies performed on the MOF MIL-125 with \ch{CH3}, \ch{NO2}, \ch{NH2}, and OH functionalizations~\cite{hend+13jacs,li+18aipa} using the range-separated hybrid functional HSE~\cite{heyd+03jcp}.
In light of the relatively large band-gap values delivered by our PBE calculations, we can speculate that most of the MOFs considered in this work are unlikely good absorbers of visible light: their absorption onset is expected to be in the ultraviolet region. 
Only the systems including OH, \ch{NH2}, and Br terminations can be possibly excluded from this estimation.
To confirm or disprove this speculation, DFT results with more advanced approximations for the exchange-correlation potential or even many-body perturbation theory calculations~\cite{kshi+21jpcl} are demanded for future work.

\begin{figure}
    \centering
    \includegraphics[width=\textwidth]{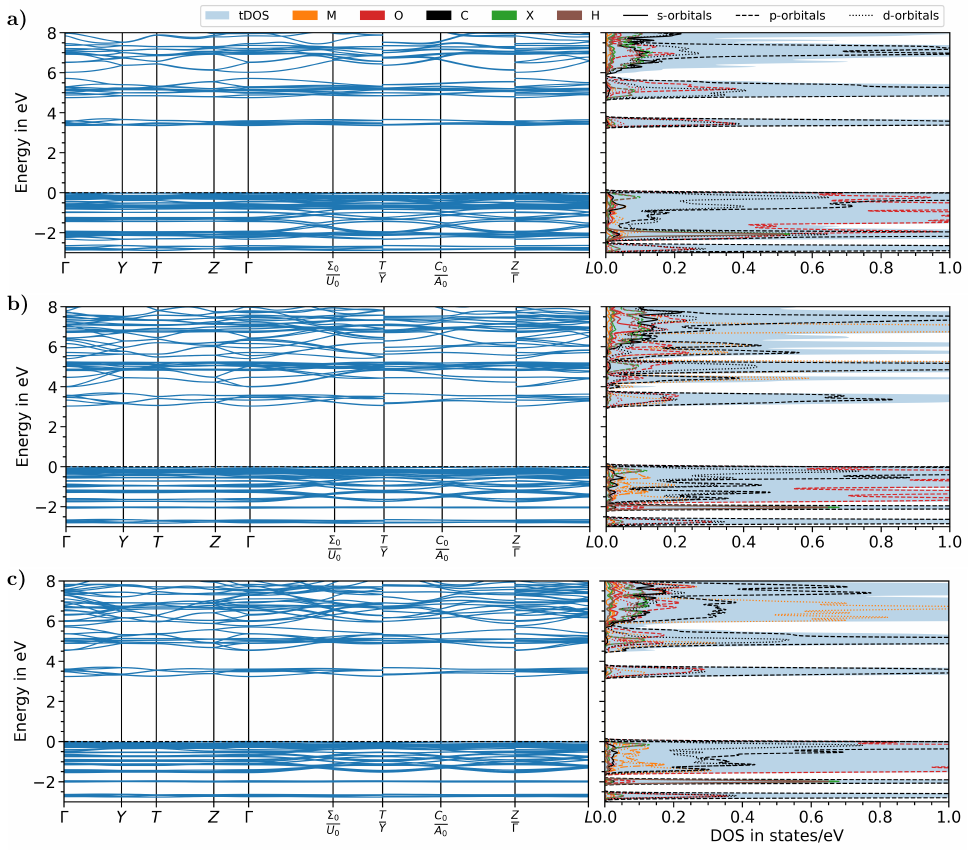}
    \caption{Band structure and projected density of states (pDOS) of the H-terminated MOFs with metal nodes a) Al, b) Sc, and c) Y. The valence band maximum is set to 0~eV. The pDOS includes contributions summed over the atomic orbitals of the metal node (M) and of the O, C, and H atoms. X labels for the H atoms passivating the phenyl ring in the sites assigned to the functional group.}
    \label{fgr:bands_H}
\end{figure}

We continue our analysis of the electronic properties of the considered MOFs by inspecting their band structures and projected density of states (pDOS).
We start from the systems with H-terminated ligands in order to rationalize first the core features of the MOFs and the impact of metal-node exchange.
As shown in Figure~\ref{fgr:bands_H}, the band gap is direct in all systems and located at the $\Gamma$-point. 
Overall, the displayed bands show little dispersion due to the strong localization of the electronic states on the organic linkers.
To corroborate this interpretation, we recall that similar features are also present in the band structures of molecular crystals and aggregates~\cite{tiag+03prb,ambr+09njp,cocc-drax17jpcm,guer+19jpcc,davi+20cm,cocc+22jpm}.
In the valence region, bands are almost flat along the $\Gamma$-Y-T-Z-$\Gamma$ path in the Brillouin zone (see Figure~S1) while larger dispersion appears elsewhere.
We can relate the band dispersion with the orientation of the linker molecules in the unit cell: along the (reciprocal) directions parallel to the carbon conjugation of the linker, the electron mobility is larger, and, hence, the bands are more dispersive.
This rationale is supported by the analysis of the pDOS which reveals a correspondence between bands characterized by a large dispersion and states dominated by C $p$-orbitals, especially in the conduction region (see Figure~\ref{fgr:bands_H}).

The electronic levels closest to the frontier exhibit carbon-oxygen hybridization, indicating the contribution of the \ch{CO2} at the connection between the BDC linkers and the metal nodes, as well as contributions from the $d$-orbitals of the metal atoms particularly for the structures with Sc and Y (Figure~\ref{fgr:bands_H}b,c).
In the case of the Al-based MOF, the orbital contributions of the metal ions are further away from the frontier (Figure~\ref{fgr:bands_H}a), due to the small size of this atomic species.
Other differences among the results shown in Figures~\ref{fgr:bands_H}a-c concern the energetic separation between the lowest manifold of conduction states which is lower in the Sc-based MOF in (panel b) compared to the other two cases: this behavior can be again ascribed to the energy of the Sc $d$-orbitals which are found in the above-mentioned region.
Moreover, in the valence, two manifolds of flat bands due to orbital contributions of Sc and Y are identified in the corresponding pDOS (see Figures~\ref{fgr:bands_H}b,c) but they are absent in Figure~\ref{fgr:bands_H}a), where Al states participate in deeper levels, as mentioned above.

\begin{figure}
    \centering
    \includegraphics[width=\textwidth]{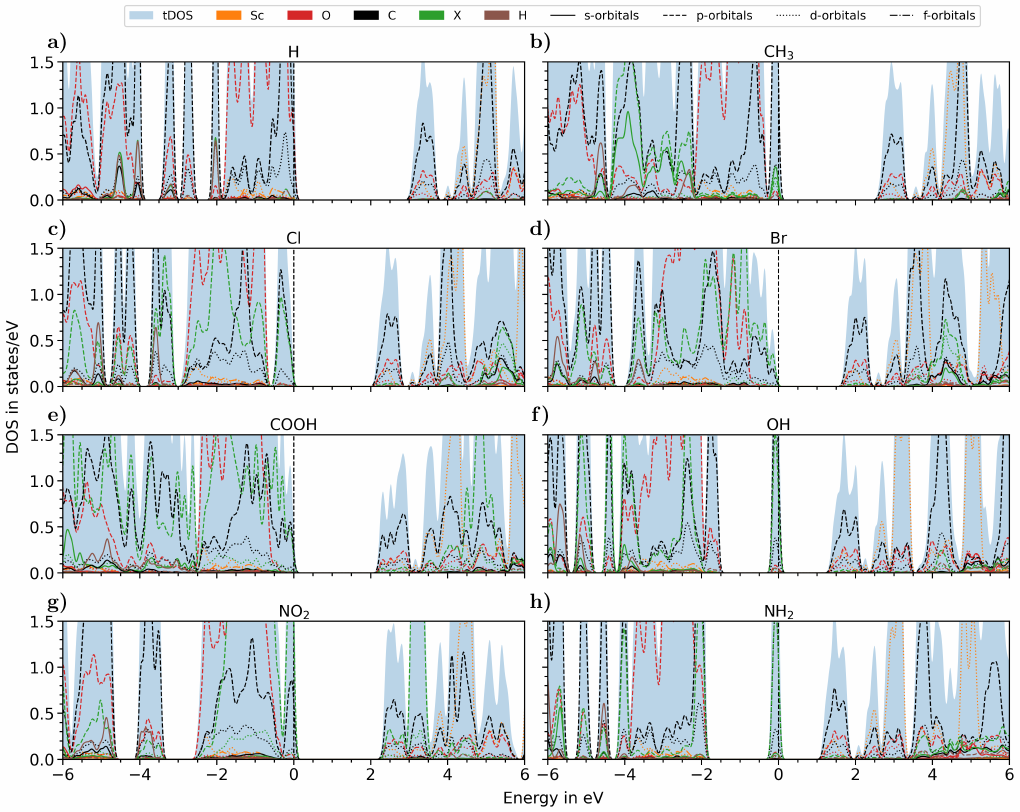}
    \caption{Projected density of states of the MOF structures with Sc nodes and varying ligand functionalizations: a) H, b) \ch{CH3}, c) Cl, d) Br, e) COOH, f) OH, g) \ch{NO2}, and h) \ch{NH2}. The valence band maximum is set to 0~eV.}
    \label{fgr:Sc_pdos}
\end{figure}
	
Supported by these findings, we continue our analysis by inspecting the impact of the linker functionalization on the electronic structure of the considered MOFs focusing on the pDOS of the Sc-based frameworks, see Figure~\ref{fgr:Sc_pdos}; the corresponding band structures are reported in Figure~S2.
At a glance, we identify the general trends for the band gaps discussed above with reference to Figure~\ref{fgr:band_gaps}.
With all terminations except for OH and \ch{NH2}, the reduction of the band gap compared to the H-passivated system is due to a rigid lowering of the lowest conduction states (with \ch{CH3}, Cl, Br, and COOH) combined with an upshift of the higher valence-band manifold (with \ch{NO2}).
In contrast, OH and \ch{NH2} terminations (Figure~\ref{fgr:Sc_pdos}f and h) give rise to a localized state with O- and N-$p$ character pinning the top of the valence band and thus effectively reducing the size of the band gap by about 50\% compared to the other groups. 
In the presence of the \ch{CH3} functionalization, the highest valence state receives contributions from the $p$- and $s$-orbitals of the methyl group (Figure~\ref{fgr:Sc_pdos}b) in addition to the $p$-orbitals of the carbon atoms of the BDC molecule characterizing its H-passivated sibling (Figure~\ref{fgr:Sc_pdos}a).
With halogen terminations, the $p$-orbitals of Cl and Br strongly contribute to the occupied frontier states owed to the large electronegativity of these elements, see Figure~\ref{fgr:Sc_pdos}c,d, leading to the discussed reduction of the gap.
We note in passing that in those systems, the valence-band maximum is shifted from the $\Gamma$ to the high-symmetry point L giving rise to an indirect band gap of 2.13 and 1.69~eV for the Cl- and Br-terminated MOFs, respectively (see Figure~S2).
Finally, with the \ch{NO2} functionalization (Figure~\ref{fgr:Sc_pdos}g), the O $p$-orbitals of the functional group contribute to the highest valence state, introducing a distinct peak at the valence-band maximum similar to the methyl group (see Figure~\ref{fgr:Sc_pdos}b).

The discussion reported above on the Sc-based MOFs can be readily extended to the systems with Al and Y nodes (see Figures~S3--S6).
As commented above with reference to Figure~\ref{fgr:bands_H}, the influence of the metal atoms on the electronic structure is expectedly small and does not affect the region around the gap. 
On the other hand, the electronic fingerprints of the ligand functionalizations, including the gap states induced by the OH and \ch{NH2} are preserved regardless of the metal node.

\section{Conclusions}

In summary, we presented a computational study based on automatized DFT calculations on the structural and electronic properties of scandium terephthalate MOFs with ligand functionalization and metal-node exchange.
In our analysis, we considered eight linker terminations, including H, \ch{CH3}, \ch{NO2}, Cl, Br, \ch{NH2}, \ch{OH}, and \ch{COOH}, as well as the isoelectronic elements Al and Y to replace Sc.
We found that all 24 considered structures are stable, exhibiting negative values of formation energies per atom.
The largest stability is obtained with COOH and OH terminations in the Sc- and Y-based MOFs.
We identified a direct correlation between the atomic radius of the metal node and the unit-cell volume.
The effects of linker termination are more pronounced in the Al-based MOFs, characterized by the smallest size compared to those with Sc and Y, and are due to the steric hindrance of the functional groups.
Structures exhibiting a larger volume are also generally more stable.
The adopted Bader charge analysis provides a metric for characterizing the bondings among the involved species. 
The coordinative character of the metal-ligand bond depends mostly on the metallic species but is modulated by the functional groups in agreement with earlier findings on zeolitic imidazolate frameworks~\cite{edza+23jpcc}.
Within the linkers, covalent bonds are formed among the C atoms of the phenyl ring except for the one bound to the functional group which acquires a positive fractional charge in the presence of the electronegative terminations OH, \ch{NO2}, and \ch{NH2}.
Furthermore, an ionic bond is formed between the carbon and oxygen atoms in the \ch{CO2} units binding the BDC molecule to the metal node.
In terms of electronic structure, all considered MOFs are semiconductors with band gaps ranging from 1.2 to 3.4~eV.
These values, obtained with the PBE functional, represent an underestimation based on which we can speculate that these systems will absorb ultraviolet radiation.
For the band gaps, we find a large influence of the functional groups, with electronegative terminations leading to about 50\% reductions compared to the H-passivated reference.
The analysis of band structures and pDOS confirm this trend and furthermore reveals the formation of gap states in the presence of OH and \ch{NH2} terminations.

In conclusion, our results demonstrate that substituting the Sc node with isoelectronic and earth-abundant elements such as Y and Al alters the volume of the MOFs but does not significantly affect their electronic properties.
In contrast, ligand functionalization has a large impact on the stability, charge distribution, and electronic structure of the systems. 
Based on these findings we can conclude that exchanging Sc atoms with Al or Y represents a sustainable alternative to preserve the fundamental characteristics of terephthalate frameworks. 
Additional tuning to target desired applications can be realized by choosing appropriate ligand functionalizations.

	\begin{acknowledgement}
This work was funded by the German Federal Ministry of Education and Research (Professorinnenprogramm III), and by the State of Lower Saxony (Professorinnen für Niedersachsen, SMART, and DyNano). J.E. appreciates additional financial support from the Nagelschneider Stiftung. 
The computational resources were provided by the North-German Supercomputing Alliance (HLRN), project nic00076 and nic00084, and by the high-performance computing cluster CARL at the University of Oldenburg, funded by the German Research Foundation (Project No. INST 184/157-1 FUGG) and by the Ministry of Science and Culture of the Lower Saxony State.

	\end{acknowledgement}
	
	\begin{suppinfo}
Bond lengths and volumes; formation energies; complete list of partial charges; band structures and projected density of states of all systems.
		
	\end{suppinfo}
	
\section*{Data Availability Statement}
The data that support the findings of this study are openly available in Zenodo at DOI 10.5281/zenodo.10082632.
	
\providecommand{\latin}[1]{#1}
\makeatletter
\providecommand{\doi}
  {\begingroup\let\do\@makeother\dospecials
  \catcode`\{=1 \catcode`\}=2 \doi@aux}
\providecommand{\doi@aux}[1]{\endgroup\texttt{#1}}
\makeatother
\providecommand*\mcitethebibliography{\thebibliography}
\csname @ifundefined\endcsname{endmcitethebibliography}
  {\let\endmcitethebibliography\endthebibliography}{}

\end{document}